\documentstyle[prb,multicol,aps,epsf]{revtex}
\begin{document}
\title{MECHANISMS OF CARRIER-INDUCED FERROMAGNETISM IN DILUTED MAGNETIC
SEMICONDUCTORS}
\author{Yu.G.Semenov}
\address{Institute of Semiconductor Physics, National Academy of\\
Sciences of Ukraine, Prospect Nauki 45, Kiev 252028, Ukraine}
\author{S.M.Ryabchenko}
\address{Institute of Physics, National Academy of Sciences of\\
Ukraine, Prospect Nauki 46, Kiev 252028, Ukraine}
\date{\today}
\maketitle

\begin{abstract}
Two different approaches to the problem of carrier-induced 
ferromagnetism in the system of the disordered magnetic ions, 
one bases on  self-consistent procedure for the exchange mean
fields, other one bases on the RKKY interaction, used in present 
literature as the alternative approximations is analyzed.
Our calculations in the framework of exactly solvable model
show that two different contributions to the 
magnetic characteristics of the system represent 
these approaches. One stems from the diagonal part of 
carrier-ion exchange interaction that corresponds to 
mean field approximation. Other one stems from the off-diagonal part 
that describes the interaction between ion spins via free
carriers. These two contributions can be responsible for the different 
magnetic properties, so aforementioned approaches are
complementary, not alternative. A general approach is 
proposed and compared with different approximations to the problem
under consideration.
\end{abstract}

%\draft

%\pacs{
%75.70.Ak, % Magnetic properties of monolayers and thin films
%05.30.Fk, % Fermion systems and electron gas
%}

\begin{multicols}{2} \narrowtext

One of the fruitful ideas of microscopic theory of magnetism is to separate
the electrons with non-compensated spins into those localized in the
crystalline lattice cites and delocalized (band) ones. The exchange
interaction between localized and delocalized electrons is then represented
as a spin density of band electrons at the cites occupied by magnetic ions.
Vonsovskii realized this idea in \cite{V1} and calculated band electron
energy shift due to $s-d$-exchange
interaction. The Hamiltonian of $s-d$-exchange
interaction was further used to study the
ferromagnetic metals (\cite{Ka}, \cite{Ko}), superfine interactions in the
solid
solutions of magnetic metals (\cite{Wa}), and magnetic impurities in
nonmagnetic metals (\cite{An}). The detailed review of this material has been
made in \cite{KPZ}. Let us note some specific features of the carrier-ion
exchange interaction. Its diagonal part is represented by the exchange
integral $J({\vec k},{\vec k}^{\prime}= {\vec k})$ over the Bloch functions
with wave vectors $\vec k$, ${\vec k}^{\prime}= {\vec k}$ and describes
so-called magnetizing effect (\cite{V1}) (or effect of redistribution
of electronic population (repopulation effect), \cite{Ze}). The latter effect
is indeed the variation of total spin of the band carriers in the effective
exchange field ${\vec G}_L$ created by localized spin moments (LSM). The
value of this variation is defined by Pauli susceptibility with field ${\vec %
G}_L$. We emphasize that the effect of giant spin splitting (GSS) of band
states (\cite{KRTZI}) discovered in diluted magnetic semiconductors (DMS) has
similar nature. From the standpoint of \cite{KRTZI}, the repopulation
effect in metals may be considered as a redistribution the band electrons
populations between spin split subbands in the same way as a single Fermi
level is established for electrons with opposite spin orientations. The
off-diagonal in $\vec k$ (\cite{Rm1}) part of the carrier-ion exchange
interactions with the $J({\vec k},{\vec k}^{\prime}= {\vec k})$ gives rise
to the spin density oscillations (\cite{Wa}). These oscillations, however,
lead to
indirect exchange interaction between the LSMs, known in metal physics as
RKKY interaction (\cite{Ru}, \cite{Ka}, \cite{Yo}).

Effective exchange field ${\vec G}_L$ is proportional to LSM magnetization ${%
\vec M}$. Latter, in turn, is defined by the sum of external magnetic field $%
{\vec B}$ and effective exchange field ${{\vec B}_e}$, which acts
on the LSM by the polarized electron spins. The LSM magnetization ${\vec M}$
also depends on spin-spin interactions between LSM. Below we will
present the arguments that both direct spin-spin interaction which
is independent on the free carriers (LL-interactions) and indirect
one induced by the free carriers (LeL -interactions) contribute to
${\vec M}$. In turn, the effective field of carriers
${{\vec B}_e}$ depends on the quantity ${\vec G}_L + g_e\mu
_B{\vec B}$, determined by electronic spin polarization, where
$g_e$ is electron g-factor, $\mu _B$ is Bohr magneton. So, the
effective exchange field ${\vec G}_L$ is expressed through itself
in a self consistent manner. Generally speaking, such self
consistency increases the magnetic susceptibility
$\chi=Const(T - \Theta)^{-1}$  (\cite{Ab}, \cite{PR}). Thus one can expect
a ferromagnetic
phase transition if $\Theta >0$. A possibility of the appearance
of dopant-induced ferromagnetism in semiconductors due to exchange
interaction of the band and localized electrons was studied for
the case of semiconductor with bivalent shallow impurities which
reveal both magnetic and electric properties (\cite{Ab}). Later
the expressions for critical temperatures of ferromagnetic
transition in the DMS with deep magnetic ion levels were obtained
in \cite{PR}. It should be
noted that self consistent procedure was essential for determination of the $%
{\vec B}_e$ and ${\vec B}_L$ exchange fields in both aforementioned works.

In recent years, it has been substantial increase of interest to
studies of the carriers induced ferromagnetism in the DMS. A
number of works (see \cite{S1} - \cite{S4} and references therein) were
devoted to the proof of the existence of ferromagnetic transition
in the DMS $Pb_{1-x-y}Sn_yMn_xTe$, induced by strong exchange
interaction of the Mn ions with the band holes with wide range (up
to $2\cdot 10^{21} cm^{-3}$) of concentrations. Ferromagnetism of
the Mn ions was found to be due to their interaction with band
holes confined in two-dimensional quantum wells on the base of DMS
$Cd_{1-x}Mn_xTe$ (\cite{D1}). The carrier-induced ferromagnetism
was also observed in the structures $A^3_{1-x}Mn_xB^5$ with $x$
about few percents. The holes in these structures are also
associated with the Mn ions (\cite{O1}).

However, the approach used in the aforementioned works (\cite{S4},
\cite{D2}, \cite{Ni})
was different from that of \cite{Ab} and \cite{PR}. Namely, the role of the
band
carriers were reduced to induction of the RKKY interaction only. So,
self-consistent contribution of diagonal part of carrier-ion exchange
interaction was not taken into account.

Note, that authors of Ref. [D2] carried out a special analysis of
magnetic susceptibility peculiarities in the assumption that role of
carrier-ion interaction is reduced to the repopulation effect only. In the
spirit of the works \cite{Ab} and \cite{PR}, they used the self consistent
procedure
for the exchange fields ${\vec G}_L$ and ${\vec B}_e$. The ferromagnetic phase
transition temperature $\Theta _{MF}$ , obtained by this procedure was shown
to coincide with transition temperature $\Theta _{RKKY}$ , calculated with
the help of RKKY interaction, considered as the sole reason for
carrier-induced ferromagnetism. Such coincidence can be obtained only under
following additional assumptions: (i) RKKY interaction can be represented in
terms of Curie-Weiss field, (ii) the magnetic ions spatial distribution
corresponds to the ideal gas of particles. The obtained coincidence $\Theta
_{RKKY}= \Theta _{MF}$ may cause an illusion of equivalence and
interchangeability of these two approaches.

It already follows from aforementioned discussion that the equality $\Theta
_{RKKY}= \Theta _{MF}$ is not strictly asserted for real systems because at
least the spatial distribution of magnetic ions corresponds to the lattice
gas rather then to ideal gas. Moreover, it is possible to imagine a
situation when the non random spatial distribution results in
antiferromagnetic transition due to oscillating nature of RKKY interaction
with $\Theta _{RKKY}<0$ while inequality $\Theta _{MF}>0$ is true for any
spatial distribution of the magnetic ions. Formally speaking, the equality $%
\Theta _{RKKY}= \Theta _{MF}$ by itself cannot be used as an evidence of the
statement that either self-consistent exchange fields consideration or
carriers-induced LeL interaction are the equivalent descriptions of the same
interactions. This stems from the aforementioned fact that they are
described by different parts of Hamiltonian. Thus, both
self-consistent mean exchange fields, leading to the repopulation effect,
and Curie-Weiss field stemming from the carriers-induced spin-spin
interaction, have to be taken into account simultaneously (\cite{Rm3}). They
are complementary to each other in the analysis of both carrier-induced
ferromagnetism in DMS and dopant-induced ferromagnetism in metals (see
\cite{V2}
and a remark to this point in \cite{PR}). In spite of the fact, that
latter conclusion follows both from the analysis of classical works
\cite{V1}, \cite{Ru}
and from detailed discussion of the differences between manifestations of
diagonal and off-diagonal parts of interaction in \cite{Wa}, it is not
completely clear if it is applicable to the case of carrier-induced
ferromagnetism in DMS. In our view this question can be clarified by the
calculation of phase transition temperature $T_c$ from the first principles
by means of exactly solvable model. In this case, we avoid the auxiliary
models for self-consistent fields ${\vec G}_L$, ${\vec B}_e$ or Curie-Weiss
field of LeL-interacting LSM. We also avoid the discussion of the
equivalence or complimentarity of the approaches under consideration.

It is impossible to calculate exact value of $T_{c}$ in real experimental
cases \cite{S1}, \cite{D1}, \cite{O1}. However, this
is not necessary since our aim is to clarify the relative role of
aforementioned mechanisms of magnetic phase transitions. Below we
give an illustrative example of exact solution for the magnetic
phase transition temperature $T_{c}$ . This example allows also to
separate contributions from self-consistent fields ${\vec{G}}_{L}$
and ${\vec{B}}_{e}$ and carriers-induced LeL spin-spin
interaction.

The Hamiltonian of our model is similar to that applied in aforementioned
works and comprises a sum of LSM , electrons and their interaction
Hamiltonians:

\begin{equation}
{\cal H}={\cal H}_{m}+{\cal H}_{e}+{\cal H}_{em},  \label{eq1}
\end{equation}
where
\begin{eqnarray*}
{\cal H}_{m} &=&g_{m}\mu _{B}B\sum_{j}S_{Z}^{j}\equiv \omega _{m}M_{L}, \\
{\cal H}_{e} &=&\sum_{b,k,\sigma }\left( \varepsilon _{b}+\omega _{e}\sigma
\right) a\dagger _{b,k,\sigma }a_{b,k,\sigma }, \\
{\cal H}_{em} &=&-\frac{J}{N_{0}}\sum_{b,b^{\prime },k,\sigma
}A_{b,b^{\prime }}\left( \sigma M\right) a\dagger _{b,k,\sigma }a_{b^{\prime
},k,\sigma }.
\end{eqnarray*}

Here $S_{Z}^{j}$ is $Z$-component of $j$-th LSM spin while $M_{L}=$ $%
\sum_{j}S_{Z}^{j}$, $j=1...N_{m}$, $N_{m}$ is the number of LSM in the
system, $g_{m}$ is the LSM g-factor, $\omega _{m}=$ $g_{m}\mu _{B}B$ is LSM
Zeeman splitting in the field $B$ and $\omega _{e}$ is delocalized electrons
Zeeman splitting in the field $B$. Three quantum numbers can be attributed
to electrons: band number $b$, intraband quantum number $k$ and projection
of spin $\sigma $ =$\pm 1/2$; $a\dagger _{b,k,\sigma }$and $a_{b,k,\sigma }$
are the creation and annihilation operators; $J$ is a constant of
carrier-ion exchange interaction; normalization factor $N_{0}$ corresponds
to one half of the number of electronic states in each of bands $b$; $%
A_{b,b^{\prime }}\left( \sigma M\right) $ is an interband transition matrix
element.

The structure of Hamiltonian (\ref{eq1}) is similar to that in \cite{Ab}-
\cite{Ni}. The difference is both in the dispersions of band carriers, $%
\varepsilon _{b,k}=\varepsilon _{b}$, that formally corresponds to flat
bands, and in the lack of intraband exchange scattering. The exchange
scattering between bands $b$ and $b^{\prime }$ with electron spin flip is
taken into account by the matrix element $A_{b,b^{\prime }}$. If we restrict
ourselves to only two electronic bands $b=1,2$, the diagonalization of the
Hamiltonian becomes trivial. Eigenenergy $E$ is defined by the
redistribution of electrons (with spins projections $\sigma $) within bands $%
b=1$ and 2 as well as by the magnetic ions (with spin projections $S_{Z}^{j}$
) distribution or, more precisely, by normalized amount of magnetic ions $%
\mu =M_{L}/N_{m}$. For simplicity we assume $A_{b,b^{\prime }}=1$. Thus, the
energy of unit volume reads:
\begin{mathletters}
\begin{eqnarray}
E_{b} &=&\frac{1}{2}n_{b}\Delta E+\left( G_{L}+\omega _{e}\right) \left(
n_{b^{+}}+n_{b^{-}}\right) \pm  \nonumber \\
&&\frac{1}{2}n_{b}\Delta E\sqrt{1+\left( \frac{G_{L}}{\Delta E}\right) ^{2}}%
+\omega _{m}n_{m}\mu .  \label{eq2}
\end{eqnarray}

The signs ''-'' or ''+'' before the square root sign in the equation (\ref
{eq2}) correspond to $b=1$ or $b=2$, $\Delta E$ is an energy interval
between these bands, the $n_{b^{+}}$ and $n_{b^{-}}$ corresponds to
concentration of electrons with the spin projection $\sigma =+1/2$ and $-1/2$
in the band $b$, the total concentration is $n_{b}=$ $n_{b^{+}}+n_{b^{-}}$, $%
G_{L}=Jx\mu $, $x=N_{m}/N_{0}$ being a fraction of magnetic cations in the
crystal, $N_{m}$ is their concentration.

Since value of $G_{L}$ is
infinitesimal at $T>T_{c}$, the square root in expression (\ref{eq2}) can be
expanded over the small parameter $(G_{L}/\Delta E)^{2}$ up to first
nonvanishing term. This term is proportional to $\mu
^{2}=(N_{m})^{-2}\sum_{j,j^{\prime }}S_{Z}^{j}S_{Z}^{j^{\prime }}$ .
Therefore, it can be considered as a contribution to the energy from
LeL spin-spin interaction induced by the band electrons. We will
also assume, that the only lowest energy band $b=1$ is filled , i.e. $E\gg kT
$. Then, the energy spectrum assumes following form:
\begin{equation}
E=n_{e}Jx\mu \sigma _{e}-n_{e}\left( \frac{G_{L}}{2\Delta E}\right)
^{2}+\omega _{m}n_{m}\mu .  \label{eq2b}
\end{equation}
For brevity, we introduce a total concentration of electrons $n_{e}=n_{b=1}$
and an average projection of electronic spins $\sigma
_{e}=(n^{+}-n^{-})/(n^{+}+n^{-})$. The first term in Eq.(\ref{eq2b})
corresponds to diagonal part of interaction (\ref{eq1}) while the second one
represents the contribution from off-diagonal (with respect to quantum
number $b$) part; the electronic $g$- factor is assumed to be equal to zero.

For magnetic susceptibility calculations we need the partition function.
This function has the form
\end{mathletters}
\begin{equation}
Z=\int \int U_{N_{m}}\left( M_{L}\right) U_{N_{e}}\left( M_{B}\right) \exp
\left( -\frac{E}{kT}\right) dM_{L}dM_{B}  \label{eq3}
\end{equation}
and can be immediately calculated with the help of Eq. (\ref{eq2b}). The
full projections (ne yasno) of LSM $M_{L}=N_{m}\mu $ and the band electrons $%
M_{B}=N_{e}\sigma _{e}$ are introduced in Eq. (\ref{eq3}). Beyond magnetic
saturation, the statistical weights $U_{N}(M)$ are defined by Gaussian
distribution in thermodynamic limit ($N_{m}\rightarrow \infty
,N_{e}\rightarrow \infty $ ) (\cite{RS}):
\begin{equation}
U_{N}(M)=\frac{\left( 2S+1\right) ^{N}}{\sqrt{\pi \Delta _{S}}}\exp \left( -%
\frac{M^{2}}{\Delta _{S}}\right) ,  \label{eq4}
\end{equation}
where $\Delta _{S}=$(2/3)S(S+1)N, S is the LSM value. Eq.(\ref{eq4}) is also
applicable for band electrons with S=1/2 if electrons obey a Boltzmann
statistics. Such approach is evidently realized in the limit $N_{e}\ll N_{0}$%
. Thus, the partition function $Z$ (and hence the magnetic susceptibility )
is calculated by straightforward integration in Eq. (\ref{eq3}) with
aforementioned assumptions. After some algebra we arrive at following final
result:

\begin{equation}
\chi _{0}=\frac{\chi _{0,L}}{1-\frac{2}{3}S(S+1)\frac{J^{2}x^{2}}{4T\Delta E}%
\frac{N_{e}}{N_{m}}-\frac{2}{3}S(S+1)\frac{J^{2}x^{2}}{8T^{2}}\frac{N_{e}}{%
N_{m}}},  \label{eq5}
\end{equation}
where $\chi _{0,L}=\frac{2}{3}S(S+1)\frac{n_{m}}{T}$ is a paramagnetic
susceptibility of noninteracting LSSM with concentration $n_{m}$

Expression (\ref{eq5}) allows an easy interpretation. In the case of band
electrons absence, $N_{e}=0$, the $\chi _{0}$ describes an ideal paramagnet.
If the electrons are present, the interband exchange scattering influence
can be excluded from consideration as $\Delta E\rightarrow \infty $ .
Clearly, this corresponds to taking into account the self-consistent
exchange fields $G_{L}$ and $B_{e}$ only. General Eq. (\ref{eq5}) gives for
this case
\begin{equation}
T_{c}=\Theta _{MF}=\sqrt{\frac{1}{12}S(S+1)J^{2}\Omega _{0}^{2}n_{m}n_{e}},
\label{eq6}
\end{equation}
where $\Omega _{0}$ is a crystal unit cell volume. It is interesting to note
that in spite of the extreme simplicity of the model under consideration,
Eq. (\ref{eq6}) reproduces a result of \cite{PR} obtained in terms of
self-consistent exchange fields for more realistic situation. It allows to
consider the third term in the denominator of Eq. (\ref{eq5}) as a
contribution from self-consistent exchange fields ${\vec{G}}_{e}$ and ${\vec{%
G}}_{L}$.

As it was mentioned above, the second term is due to off-diagonal part of
carrier-ion exchange Hamiltonian. With respect to this part only, we have to
omit the last term in denominators of Eq.(\ref{eq5}). In this case we obtain
\begin{equation}
T_{c}=\Theta _{ind}\equiv \frac{1}{6}S(S+1)\frac{J^{2}\Omega
_{0}^{2}n_{m}n_{e}}{\Delta E}.  \label{eq7}
\end{equation}

It is seen that Eq. (\ref{eq7}) is not similar to Eq. (\ref{eq6}), neither
quantitatively nor qualitatively. It gives possibility to make a following
general statement. For the problems of magnetic phase transitions, as well
as for the calculations of magnetic susceptibility, magnetization, etc. it
is important to simultaneously take into account both parts (diagonal and
off-diagonal) of carrier-ion exchange interaction. Therefore, the omission
of any one of the aforementioned terms in the Hamiltonian leads, generally
speaking, to significant inaccuracy or even to qualitative changes.

We shall present now the way of consideration of the problem of
spontaneous magnetic transitions induced by band carriers in DMS,
which we consider to be correct. We choose the Hamiltonian in the
form (1), but will incorporate there the LL spin-spin interaction
between LSM ${\cal H}_{LL}$ (\cite{PR}) to the ${\cal H}_{m}$  and
the intraband exchange scattering between Bloch electron states
(\cite{SSh}).

To calculate the magnetic susceptibility with the help of
modified Hamiltonian (1), we shall carry out the approximate
diagonalization of Eq. (1) by elimination of its off-diagonal (in
${\vec k}$ and ${\vec k}^{\prime }$ ) components by canonical
transformation method (\cite{BP}) in second order of perturbation
theory. As a result, the operator of effective LeL spin-spin
interaction assumes the form:

\begin{equation}
{\cal H}_{LeL}=\sum_{j,j^{\prime }}J_{eff}\left( {\vec
R}_{j,j^{\prime }}\right) {\vec S}^{j}{\vec S}^{j^{\prime }},
\label{eq8}
\end{equation}
where ${\vec R}_{j,j^{\prime }}$ is a radius-vector joining the
pairs of magnetic
ions in the crystal lattice sites $j$ and $j^{\prime }$. The structure of $%
{\cal H}_{LeL}$ is similar to the ${\cal H}_{LL}$ in magnetic
Hamiltonian ${\cal H}_{m}$ and can be added to it. The specific
form of $J_{eff}\left( {\vec R}_{j,j^{\prime }}\right)$ is defined
by electronic gas degeneration (\cite{Kl}), influence of magnetic
field (\cite{Kh}), effect to casual anisotropy (Ref. [IK]),
structure of energy band of semiconductor (\cite{S4}) and
dimensionality of the system (\cite{D2}, \cite{Ya}).

The diagonal part of the operator ${\cal H}_{em}$ can be written
down in the form of LSM's Zeeman energy in the effective field $%
B_{e}=J\Omega _{0}n_{e}\sigma _{e}/g_{m}\mu _{B}$. This part can be added to
the Zeeman term of magnetic Hamiltonian ${\cal H}_{m}$. One more
standard step is the transformation of spin-spin interactions in ${\cal H}%
_{m}$ and ${\cal H}_{LeL}$ (\ref{eq8}) to Curie-Weiss fields. Such
approach, as it is well known, reduces the thermodynamical treatment of
interacting spin to the consideration of isolated spins with the
effective temperature $T_{eff}=T-\Theta $. The parameter $\Theta =\Theta
_{LL}+\Theta _{LeL}$ is defined by both LL interaction and 
LeL interaction (Eq.(\ref{eq8})).

As a result the free energy can be presented in terms of electronic and
ionic parts only (\cite{SS}):

\begin{equation}
F=F_{e}\left( \sigma _{e}\right) +F_{m}\left( B+B_{e},\ T-\Theta \right) ,
\label{eq9}
\end{equation}
where $F_{m}\left( B+B_{e},\ T-\Theta \right)$ is a contribution
of noninteracting (isolated) spins subjected to the uniform
magnetic field $B+B_{e}$ at the temperature $T-\Theta$. Note that
the Eq. (\ref{eq9}) takes into account both diagonal part of
carrier-ion exchange interaction (term $B_{e}$) and its
off-diagonal part (term $\Theta _{ind}$). The electronic
polarization $\sigma _{e}$ is calculated by minimization of
functional (\ref{eq9}). Further substitution of $\sigma _{e}$
obtained in such manner to Eq. (\ref{eq9}) defines completely
thermodynamic characteristics of the system: magnetization
$M_{\alpha }=-\partial F/\partial B_{\alpha }$,
susceptibility$\chi _{\alpha \beta}=$ $-\partial ^{2}F/\partial B
_{\alpha}\partial B_{\beta } $, $\alpha ,\beta=x,y,z$ and the
temperature of magnetic phase transition.

Specific form of free energy functional (\ref{eq9}) depends on
aforementioned and many other peculiarities of our system. As an
illustration, we consider now the most popular case of degenerated
electronic gas in a simple isotropic band of semiconductor. We consider the
magnetic transition temperature $T_{c}$ based on the preceding results
(\cite{Ab}, \cite{PR}, \cite{S1}, \cite{S4}, \cite{D2}, \cite{V2},
\cite{SS}). The Eq. (\ref{eq9}) permits to obtain
following equation for the critical temperature point:

\begin{equation}
\left( T_{eff}\right) _{c}-\Theta _{MF}=0.  \label{eq10}
\end{equation}

Here, $\Theta _{MF}$ is defined by corresponding formulas of the \cite{PR},
\cite{D2}, where only the diagonal part of the interaction operator ${\cal H}%
_{em}^{\prime }$ is taken into account and $\left( T_{eff}\right)
_{c}=T_{c}-\Theta _{LL}-\Theta _{RKKY}$. Parameter $\Theta _{RKKY}$
coincides with the $\Theta _{MF}$ only under the conditions, mentioned in
the introduction to this paper(\cite{D2}). Parameter $\Theta _{LL}$ should
be taken
from the experiment, $\Theta _{LL}=-T_{0}$ , where $T_{0}>0$ corresponds to
antiferromagnetic LL exchange interaction realized in majority of
experimental situations for DMS (Ref. [15,19,20]). So, for $T_{c}$ we can
obtain $T_{c}=2\Theta _{MF}-T_{0}$. If one takes into account only
self-consistent exchange mean fields or RKKY interactions, the
value of $T_{c}$ is determined by other expression: $T_{c}=\Theta _{MF}-T_{0}
$. This difference can be important to the prediction of conditions for
carriers-induced ferromagnetism realization in different experimental
situations.

We had shown that neglecting of any part of interactions ${\cal H}_{em}$
leads to essential reduction of predicted $T_{c}$ value even in the simplest
models. Moreover, considered example shows that such neglecting leads to the
qualitatively different results.
Present work shows that both diagonal and off-diagonal in ${\vec k}$ parts of
carrier-ion exchange interaction are important. Their simultaneous
consideration is shown to change the part of earlier results quantitatively
or even qualitatively. Nevertheless, main conclusion
of preceding works remains valid: carrier induced ferromagnetic transition
in DMS is possible under high enough carriers concentrations and reduction
of the system dimensionality enhances this effect.

Let us finally note that we have considered the necessity of simultaneous
consideration of both contributions of carrier-ion exchange interaction. One
can see that effect is essential. It is clear, that similar approach have to
be applied to many other cases. For instance, it would be useful to take
into account the LeL exchange interaction of the LSM via conduction
electrons in the problem of free or bound magnetic polaron.

\end{multicols}


\begin{references}
\bibitem{Rm3}  Not for all situations, however, this is plainly formulated.
Sometimes, consideration of no more than one part of a carrier-ion exchange
interaction (\cite{Ab}-\cite{Ni}) leads to qualitatively correct but not
quantitatively exact results. In a number of cases, the corresponding
corrections are important.
\end{references}
\end{document}